\documentclass[twocolumn,english,aps,prl,showpacs,superscriptaddress]{revtex4}
\usepackage{amsmath}
\usepackage{graphicx}
\usepackage{esint}

\providecommand{\LyX}{L\kern-.1667em\lower.25em\hbox{Y}\kern-.125emX\@}
\providecommand{\tabularnewline}{\\}

\begin{document}

\title{Collective Uncertainty Entanglement Test}

\author{\L{}ukasz Rudnicki}

\email{rudnicki@cft.edu.pl}

\affiliation{Center for Theoretical Physics, Polish Academy of Sciences, Aleja Lotnik{\'o}w 32/46, PL-02-668 Warsaw, Poland}

\author{Pawe{\l} Horodecki}
\affiliation{Faculty of Applied Physics and Mathematics, Technical University of Gda{\'n}sk, PL-80-952 Gda{\'n}sk, Poland}
\affiliation{National Quantum Information Centre of Gda{\'n}sk, PL-81-824 Sopot, Poland}
\author{Karol \.{Z}yczkowski}

\affiliation{Center for Theoretical Physics, Polish Academy of Sciences, Aleja Lotnik{\'o}w 32/46, PL-02-668 Warsaw, Poland}
\affiliation{Smoluchowski Institute of Physics, Jagiellonian University, ul. Reymonta 4, PL-30-059 Krak{\'o}w, Poland}

\begin{abstract}
For a given pure state of a composite quantum system
we analyze the product of its projections onto
a set of locally orthogonal separable pure states.
We derive a bound for this product
analogous to the entropic uncertainty relations.
For bipartite systems the bound is saturated for
maximally entangled states and it allows us
to construct a family of entanglement measures, we shall call {\sl collectibility}.
As these quantities are experimentally accessible,
the approach advocated contributes to the task
of experimental quantification of quantum entanglement,
while for a three--qubit system it is capable
to identify the genuine three-party entanglement. 

\end{abstract}
\maketitle

The phenomenon of quantum entanglement -- non--classical correlations between individual subsystems --
is a subject of an intense research interest \cite{MCKB05,BZ06,HHHH09}.
Several criteria of detecting entanglement are known \cite{BZ06,HHHH09},
and some of them can be implemented experimentally (see \cite{PR_Guhne}
for the review of specific experimental schemes). In particular 
the issue of qualitative entanglement detection is quite well established
including entanglement witnesses method (see \cite{HHHH09}) and 
local uncertainty relations \cite{Uncertainty}.
On the other hand, although various measures of quantum entanglement
are analyzed 
\cite{PV07,HHHH09}, in general they are more difficult to be quantitatively measured 
in a physical experiment.
To estimate experimentally the degree of entanglement of a given quantum state
one usually relies \cite{SM+08} on quantum tomography or analogous techniques. 

The idea of entanglement detection and estimation without prior tomography 
\cite{HE02,PH03} involves the collective measurement of two (or more) 
copies of the state as demonstrated in \cite{BCEHAS05}.
Consequently recent attempts towards experimental quantification of entanglement are 
based on finding collectively measurable quantities which bound known entanglement measures 
from below and are experimentally accessible \cite{MKB05a,WRDMB06} (for review see \cite{AL09}).

The main aim of this work is to construct 
a family of indicators, designed to quantify the
entanglement of a pure state of an arbitrary composite system,
which can be measured in a coincidence experiment
without attempting for a complete reconstruction of the quantum state.

Our approach, which leads to simple collective entanglement test, is inspired by the entropic uncertainty
relations which are satisfied by any pure state. For instance,
the sum of the Shannon entropies of the expansion coefficients
of a given pure state $|\psi\rangle \in {\cal H}_N$ expanded in two mutually unbiased bases
is bounded from below by $\ln N$ \cite{MU88}.
This observation suggests to quantify the pure states entanglement
by a function of the projections of the analyzed state $|\Psi\rangle$
of a composite system onto mutually orthogonal separable pure states. 

The method we propose can be formulated in a rather general case
of a normalized pure state, $\langle \Psi | \Psi\rangle =1$,
of a composite system consisting of $K$ subsystems.
For simplicity we shall assume here that all their dimensions are equal,
so we consider an element of a \textit{K-partite Hilbert space} $\mathcal{H}=\mathcal{H}^{A}\otimes {H}^{B} \ldots\otimes\mathcal{H}^{K}$,
where $\dim\left(\mathcal{H}^{A}\right)=\ldots=\dim\left(\mathcal{H}^{K}\right)=N$.
Let us select a set of $N$ separable pure states of a $K$--quNit system,
 $|\chi_{j}^{sep} \rangle = |a_{j}^{A} \rangle
    \otimes\ldots\otimes\left|a_{j}^{K}\right\rangle $.
where $|a_{j}^{I} \rangle\in \mathcal{H}^{I}$ with $j=1,\dots,N$ and $I=A,\dots,K$.
The key assumption is that all local states are mutually orthogonal,
so that 
\begin{equation}
\left|a_{1}^{I}\right\rangle ,\ldots,\left|a_{N}^{I}\right\rangle 
\in\mathcal{H}^{I},\qquad\left.
\left\langle a_{j}^{I}\right|a_{k}^{I}\right\rangle =\delta_{jk}.
\label{ortho}
\end{equation}

\paragraph{Entanglement detection ---}
In order to construct measurable indicators of quantum entanglement
and find practical entanglement criteria valid 
for any analyzed state $| \Psi\rangle$ we define now the following quantity 
\begin{equation}
Y^{max}\left[\left|\Psi\right\rangle \right]=\max_{\left|\chi^{sep}\right\rangle }\prod_{j=1}^{N}\left|\!\left.
\left\langle \Psi\right|\chi_{j}^{sep}\right\rangle \right|^{2}.
\label{ampli}
\end{equation}
This product of the projections of the state
onto the set of $N$ separable states, optimized 
over all possible sets of mutually locally orthogonal states,
 $|\chi^{sep}\rangle = \{ |\chi_{1}^{sep} \rangle,
 \ldots,  | \chi_{N}^{sep} \rangle \}$,
will be called maximal {\sl collectibility}.


Note the difference with respect to the geometric measure
of entanglement \cite{WG03}, to define which one takes the maximum
over a {\sl single} separable state, $|\chi_{1}^{sep} \rangle$.
In this case this maximum, denoted in \cite{WG03} by $\Lambda^2_{max}$, is equal to unity if the analyzed state $|\Psi\rangle$
is separable and it is smaller for any entangled state,
so to define the geometric measure of entanglement one takes $1-\Lambda^2_{max}$.
In contrast, taking in (\ref{ampli}) the maximum of the
product of the projections of $|\Psi\rangle$ onto $N \ge 2$ separable states
 $|\chi_{j}^{sep} \rangle$  we face an inverse situation:
we show below that $Y^{max}$ is the largest for maximally
entangled states, so this quantity can serve directly
as a quantificator of entanglement.


To this end we shall start with a variational equation
\begin{equation}
\frac{\delta}{\delta\left|\Psi\right\rangle }\left(\prod_{j=1}^{N}\left|\!\left.
\left\langle \Psi\right|\chi_{j}^{sep}\right\rangle \right|^{2}-\lambda\!\left.\left\langle \Psi\right|\Psi\right\rangle \right)=0,
\label{variation}
\end{equation}
where $\lambda$ plays the role of a Lagrange multiplier associated
with the normalization constraint.
This idea was developed by Deutsch
in order to obtain the entropic uncertainty relation \cite{Deutsch}.
Equation (\ref{variation}) implies 
\begin{equation}
\prod_{j=1}^{N}\left|\!\left.
\left\langle \Psi\right|\chi_{j}^{sep}\right\rangle \right|^{2}\sum_{i=1}^{N}\left(\!\left.\left\langle \chi_{i}^{sep}\right|\Psi\right\rangle \right)^{-1}\left\langle \chi_{i}^{sep}\right|=\lambda\!\left\langle \Psi\right|.
\label{variation_cd}
\end{equation}
Multiplying (\ref{variation_cd}) by $\left|\Psi\right\rangle $ we
find out that $\lambda=N\cdot \prod_{j=1}^{N}\left|\!\left.
\left\langle \Psi\right|\chi_{j}^{sep}\right\rangle \right|^{2}$. Moreover, the contraction
of (\ref{variation_cd}) with $\left|\chi_{m}^{sep}\right\rangle $
leads to $\left|\!\left.\left\langle \Psi\right|\chi_{m}^{sep}\right\rangle \right|^{2}=1/N$ for all values of $m$.
From this result we have
\begin{equation}
\max_{\left|\Psi\right\rangle }\prod_{j=1}^{N}\left|\!\left.
\left\langle \Psi\right|\chi_{j}^{sep}\right\rangle \right|^{2}=\prod_{j=1}^{N}\frac{1}{N}=N^{-N},
\label{max1}
\end{equation}
which after formal optimization over ${\left|\chi^{sep}\right\rangle }$ implies the desired inequality 
\begin{equation}
Y^{\rm max}\left[\left|\Psi\right\rangle \right]\leq N^{-N}.
\label{nieoznaczonosc}
\end{equation}
Using an auxiliary variable, $Z^{\rm max}=-\ln Y^{\rm max}$
this relation takes the from  $Z^{\rm max}[\Psi] \ge N \ln N$,
analogous to the entropic uncertainty relation.
Interestingly, for a bipartite system
this inequality is saturated for the maximally entangled state,
$|\Psi_+\rangle=\frac{1}{\sqrt{N}} \sum_{i} |i,i\rangle$.
while in the case of $K$--quNit system it is saturated 
for a generalized GHZ state,
$|GHZ\rangle_K=\frac{1}{\sqrt{N}} \sum_{i} |i\rangle_A \otimes \dots \otimes |i\rangle_K$.

Consider now the other limiting case of a separable state $\left|\Psi_{\rm sep}\right\rangle=\left|\Psi_A\right\rangle\otimes\ldots\otimes\left|\Psi_K\right\rangle$.
In this case the projections factorize,
\begin{equation}
\left.\left\langle \Psi_{\rm sep}\right|\chi_{j}^{sep}\right\rangle =\prod_{I=A}^{K}\left.\left\langle \Psi_{I}\right|a_{j}^{I}\right\rangle .
\end{equation}
Furthermore, for each value of the index $\left(I=A,\dots ,K \right)$ 
 we can independently apply the result (\ref{max1}) and obtain 
\begin{equation}
\prod_{j=1}^{N}\left|\left.\left\langle
\Psi_{I}\right|a_{j}^{I}\right\rangle \right|^{2}\leq N^{-N}.
\end{equation}
Thus, for any separable state we have
\begin{equation} \label{nieonzsepgen}
\prod_{j=1}^{N}\left|\!\left.
\left\langle \Psi_{\rm sep}\right|\chi_{j}^{sep}\right\rangle \right|^{2} \leq
\prod_{I=A}^{K}\max_{\left\{ \left|\Psi_{I}\right\rangle \right\}} 
\prod_{j=1}^{N}\left|\left.
\left\langle \Psi_{I}\right|a_{j}^{I}\right\rangle \right|^{2}=N^{-N\cdot K}, 
\end{equation}
so that
\begin{equation}
Y^{max}\left[\left|\Psi_{\rm sep}\right\rangle \right]\leq N^{-N\cdot K}.
\label{nieoznsep}
\end{equation}
This observation leads to the following separability criteria
based on the maximal collectibility:
\begin{equation}
\left(Y^{max}\left[\left|\Psi\right\rangle \right]>\alpha_{K,N}\right)
\Rightarrow\left(\left|\Psi\right\rangle \textrm{- entangled}\right) .
\label{criteria}
\end{equation}
Here  $\alpha_{K,N}=N^{-N\cdot K}$ is the discrimination parameter.

\paragraph{Multi qubit systems ---}
In the definition  (\ref{ampli}) of the maximal {\sl collectibility}
one performs a maximization over the set of all $N$
mutually orthogonal separable states  $|\chi_{j}^{sep} \rangle$.
The maximal collectibility  $Y^{max}$ can be considered
as a pure state entanglement measure, and we derive below
its explicit expression in the simplest case of a  two qubit system.
However, it is also 
convenient to perform the optimization procedure stepwise
and to consider first an optimization over a single separable state.

Let us then define a one-step maximum over the separable states
belonging to the first subspace ${\cal H}^A$,
\begin{equation}
Y_a\left[\left|\Psi\right\rangle \right]=\max_{\left|a^{A}\right\rangle }\prod_{j=1}^{N}\left|\!\left.\left\langle \Psi\right|\chi_{j}^{sep}\right\rangle \right|^{2}.
\end{equation}
Note that the {\sl collectibility} $Y_a$, a function of the analyzed state $|\Psi \rangle$,
is parameterized by the set $a$ of $N$ product states 
$|a_j^{B} \rangle\otimes\ldots\otimes |a_j^{K} \rangle$, with $j=1,\ldots,N$.
By construction one has 
 $\max_{a}
 Y_a\left[\left|\Psi\right\rangle
 \right]=Y^{max}\left[\left|\Psi\right\rangle \right]$.

Consider now the case of a $K$--qubit system ($N=2$).
Writing an equation analogous to (\ref{variation}) and
following the standard variational approach 
we obtain an analytical formula for the collectibility,
\begin{equation}
Y_a\left[\left|\Psi\right\rangle \right]=\frac{1}{4}\left(\sqrt{G_{11}G_{22}}+\sqrt{G_{11}G_{22}-
\left|G_{12}\right|^{2}}\right)^{2},
\label{miara}
\end{equation}
expressed in terms of elements of the Gram matrix defined for a set of projected states.
Here $G_{jk}=\!\left.\left\langle \varphi_{j}\right|\varphi_{k}\right\rangle $,
while $|\varphi_{j} \rangle \in {\cal H}^{A}$ 
denotes the state $|\Psi\rangle$ projected onto the $j$--th  separable state 
living in $K-1$ subspaces labelled by $B,\dots K$, so that
$| \varphi_{j}\rangle = [\langle a_{j}^{B}| \otimes\ldots\otimes \langle a_{j}^{K}|
]| \Psi\rangle $.

Because of (\ref{max1}) and (\ref{nieonzsepgen}) the collectibility $Y_a$ satisfies the same 
uncertainty relations (\ref{nieoznaczonosc})
and (\ref{nieoznsep}) as the maximal collectibility $Y^{max}$. 
This approach can be generalized to the case of 
Hilbert spaces with different dimensions.
It can be especially useful when $\dim\left(\mathcal{H}^{A}\right)$
is much larger than the dimensions of remaining Hilbert spaces. This
case may for instance describe the entanglement with an \textsl{environment}.

\paragraph{Two qubits ---}
 Let us now investigate in more detail the simplest 
case of a two--qubit system
for which $K=N=2$ and $\mathcal{H}=\mathcal{H}^{A}\otimes\mathcal{H}^{B}$.
Any pure state $|\Psi\rangle_{AB}$ can be then written in its Schmidt form \cite{BZ06},
\begin{equation}
\left|\Psi\right\rangle_{AB} =\left(U_A\otimes U_B\right)\left[\cos\left(\frac{\psi}{2}\right)\left|00\right\rangle +\sin\left(\frac{\psi}{2}\right)\left|11\right\rangle\right] ,
\label{qubit}
\end{equation}
where $U_A\otimes U_B$  is a local unitary. The Schmidt angle $\psi\in\left[0,\pi\right]$
is equal to zero for the separable state and 
to $\pi/2$ for the maximally entangled state.
From the uncertainty relation (\ref{nieoznaczonosc}) we know that
$Y_a\left[\left|\Psi\right\rangle_{AB} \right]\leq1/4$. Moreover, 
if the state (\ref{qubit}) is separable we have (\ref{nieoznsep})
 $Y_a\left[\left|\Psi_{\rm sep}\right\rangle \right]\leq 1/16$.

Now we assume the general form of the 
orthonormal detector basis spanned in the second subspace
$\mathcal{H}^{B}$,
\begin{eqnarray}
\left|a_{1}^{B}\right\rangle & = & \cos\left(\frac{\theta}{2}\right)\left|0\right\rangle +e^{\imath\phi}\sin\left(\frac{\theta}{2}\right)\left|1\right\rangle , \nonumber \\
\left|a_{2}^{B}\right\rangle &=& \sin\left(\frac{\theta}{2}\right)\left|0\right\rangle -e^{\imath\phi}\cos\left(\frac{\theta}{2}\right)\left|1\right\rangle ,
\label{baza}
\end{eqnarray}
where $\theta\in\left[0,\pi\right]$ and  $\phi\in\left[0,2\pi\right]$. Due to this general form, our analysis becomes independent of the local unitary $U_B$ in (\ref{qubit}). Note also that the expression (\ref{miara}) is independent of $U_A$, 
thus our approach works universally for any two-qubit pure state.
Using (\ref{baza}) we shall calculate the entries of the Gram matrix and find
\begin{equation}
Y_{\theta}\left(\psi \right)=\!\frac{\left(2\sin\left(\psi\right)+\sqrt{3-2\cos\left(2\theta\right)\cos^{2}\left(\psi\right)\!-\cos\left(2\psi\right)}\right)}{64}^{2}\!\!\!.
\nonumber
\label{Y-dependence}
\end{equation}
The collectibility $Y_{\theta}\left(\psi \right)$ depends on the analyzed
state ($\psi$) and the detector parameters $a=(\theta,\phi)$. The dependence
on the azimuthal angle $\phi$ is trivial. If the state
(\ref{qubit}) is maximally entangled ($\psi=\pi/2$ ) then
$Y_{\theta}\left(\pi/2\right)=1/4$
and the collectibility attains its maximal possible value independently of the choice
of $\left(\theta,\phi\right)$.

\begin{figure}
\centering{}\includegraphics[scale=0.73]{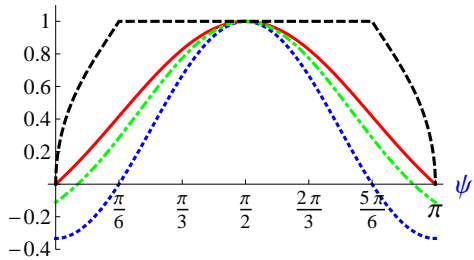}
\caption{Color online.
Parameters describing entanglement of a two--qubit
pure state $\left|\Psi\right\rangle_{AB}$ as a function of the Schmidt angle $\psi$.
We plot the minimal (blue/dotted), the average (green/dashed--dotted) and the maximal
(red) values of the rescaled collectibility
 $\left[16Y_\theta\left(\psi\right)-1\right]/3$.
Positive values identify entanglement. The black, dashed
line shows the probability $P_Y$ that the entanglement 
of $\left|\Psi\right\rangle_{AB}$ is detected in a particular random measurement.
}
\label{Fig.1}
\end{figure}

In order to characterize various possibilities to 
detect the entanglement we analyze four quantities.
Consider first the minimal ($\theta=0$) and the maximal ($\theta=\pi/2$)
value of the collectibility $Y_\theta\left(\psi\right)$ with respect to the
detector parameters  $\left(\theta,\phi\right)$,
\begin{equation}
Y^{min}\left(\psi\right)=\frac{\sin^{2}\left(\psi\right)}{4},\qquad Y^{max}\left(\psi\right)=\frac{\left(1+\sin\left(\psi\right)\right)^{2}}{16}.
\nonumber
\label{minmax}
\end{equation}
Then define the mean collectibility $\overline Y=\langle Y_a\rangle_a$, averaged over
the set of the detector parameters  $a=\left(\theta,\phi\right)$
with the measure $d\Omega=\sin\left(\theta\right)d\theta\, d\phi/\left(4\pi\right)$.
This case,  corresponding to the average over a random choice of the detector parameters, 
$\overline{Y}\left(\psi\right)=\int_{\mathcal{S}^{2}}d\Omega\, 
  Y_{\theta}\left(\psi \right)$,
yields the result
\begin{equation}
\overline{Y}\left(\psi\right)=\frac{11-7\cos\left(2\psi\right)+3(\pi - 2\psi)\tan\left(\psi\right)}{96}.
\end{equation}
Furthermore, we study the probability that the 
entanglement is detected in a measurement with a 
random choice of the detector angle $\theta$,
$P_{Y}\left(\psi\right)=\int_{\mathcal{P}}d\Omega$, where $\mathcal{P}=\left\{ \left(\theta,\phi\right)\in\mathcal{S}^{2}:\, Y_{\theta}\left(\psi\right)>1/16\right\} $:
\begin{equation}
P_{Y}\left(\psi\right)=
    \begin{cases}
\frac{\sqrt{2\sin\left(\psi\right)-\sin^{2}\left(\psi\right)}}{\left|\cos\left(\psi\right)\right|} & \textrm{ for }\psi\in\left[0,\frac{\pi}{6}\right]\cup\left[\frac{5\pi}{6},\pi\right]\\
1 & \textrm{ for }\psi\in\left[\frac{\pi}{6},\frac{5\pi}{6}\right]\end{cases}.
\label{prob}
\end{equation}

Analytical results for a pure state of the $2 \times 2 $ system 
are presented in Fig. \ref{Fig.1}.
In the case of the optimal choice of the detector parameters (red curve) the
entanglement is detected for any entangled state. More importantly,
in the case of the worst possible choice of the measurement parameters
represented by the blue/dotted curve,
the entanglement is detected for $\psi\in\left[\pi/6,5\pi/6\right]$. 
This coincides
with the fact that the probability of entanglement detection with
a single random measurement is equal to unity (cf. (\ref{prob})
and the black, dashed curve). The average collectibility $\overline Y$ 
corresponds to an average obtained by a sequence of measurements
with a random choice of the detector parameters.
\begin{figure}
\centering{}\includegraphics[scale=0.30]{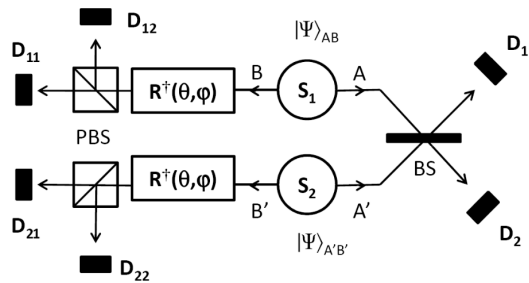}
\caption{Determination of the Gram matrix via conditional 
overlapping in the case of two polarization--entangled photon pairs.
Each source produces a pair of photons in a polarization state $\left|\Psi\right\rangle_{AB}$. 
On the left side $B$ the statistics of pairs of clicks after two PBS--elements are measured.
On the right side $A$ the Hong--Ou--Mandel interference is performed. 
The number $|G_{12}|^2$ is
equal to the probability of the pair of the clicks at $B$
multiplied by that of double click at $A$.}
\label{Fig.2}
\end{figure}
Looking at the expression (\ref{miara}) we see that 
to compute the collectibility $Y_a$ it is enough to
determine the elements of the Gram matrix.
Assume first that we analyze a two--photon polarization entangled state.
The diagonal element $G_{jj}$ 
represents an amplitude of the state $|\varphi_j\rangle$
in the first subspace ${\cal H}^A$, under the assumption that the second 
photon was measured by the detector in the state $|a^B_j\rangle$.
To determine the absolute value of the off diagonal element,  
$|G_{12}|^2=|\langle \varphi_1|\varphi_2\rangle|^2$, 
of the two--photon state $\left|\Psi\right\rangle_{AB}$
one projects the ${\cal H}^B$ part of the
first copy onto the state $|a^B_1\rangle$,
the same part of the second copy onto $|a^B_2\rangle$,
and performs a kind of the Hong--Ou--Mandel
interference experiment \cite{HOM87} with the remaining two photons
 of the first subsystem  ${\cal H}^A$. 
A specific scheme of this kind is depicted in Fig.2.

Apart from two sources of pure entanglement (which may base on 
type--I PDS sources modified by dumping one of the polarization components) it involves 
the 50:50 beamsplitter (BS), two polarization rotators $R^{\dagger}(\theta, \phi)$
in the same setting and the polarized beamsplitters (PBS). If by $p_{ij}(+,+)$ we denote 
the probability of double click after the beamsplitter, and by $p_{1i}\equiv p_{1}\left((-1)^{i+1}\right)$  ($p_{2i}\equiv p_{2}\left((-1)^{i+1}\right) $) the probability of click in the 
$D_{1,i}$-th detector ($D_{2,i}$-th detector) i.e. one of the detectors located after upper 
PBS (lower PBS) then all the Gram matrix elements are:
\begin{equation}
|G_{ij}|^{2}=p_{1i}p_{2j}(1-2p_{ij}(+,+)).
\end{equation}
Alternatively one can apply the following 
network designed to measure all three quantities (see Fig.3).
\begin{figure}
\centering{}
\includegraphics[scale=0.40]{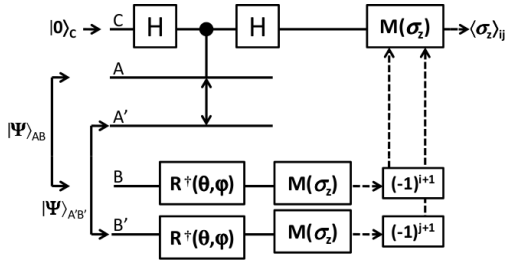}
\caption{Quantum network exploiting two copies of an analyzed state $|\Psi\rangle_{AB}$,
a control qubit $|c\rangle$ initially in state $|0\rangle$, 
 controlled SWAP gate 
(cf. \cite{HE02}) and two Hadamard gates.
The mean value of Pauli $\sigma_z$ matrix of
$|c\rangle$   
is measured
under the condition that the chosen pair $(i,j)$ of results 
is obtained in measurement of the same observables performed 
on both qubits at the bottom of the scheme.}
\label{Fig.3}
\end{figure}
Measuring the $ \sigma_z $ component of the first qubit, 
conditioned by pair of the results $(i,j)$ (coming with probabilities $p_{1}((-1)^{i+1})$, $p_{2}((-1)^{j+1})$) 
of the measurements of the same ($ \sigma_z $) observable on the last two qubits 
one gets an estimation of the parameter $|G_{ij}|^2=p_{1}((-1)^{i+1}) p_{2}((-1)^{j+1}) \langle \sigma_z \rangle_{ij}$.
Without going into detailed analysis here we only mention that
the purity assumption may be dropped at a price of performing two variants of the 
experiment each with one of two complementary (in Heisenberg sense) settings  
$R^{\dagger}(\theta, \phi)$,  $R^{\dagger}(\theta', \phi')$.  
Then the discrimination parameter $\alpha_{K,N}=\alpha_{2,2}=1/16$
in the inequality (\ref{criteria}) may be successfully corrected to take into account
noise in Hong--Ou--Mandel interference occurring in both variants. Explicit derivation of 
such correction is rather complicated and will be considered in details elsewhere. 


\paragraph{Three qubits ---} Now let us investigate the case of a
three qubit state ($K=3$). In this case the separability discrimination
parameter is equal to $\alpha_{3,2}=1/64$. We compare a bi--separable state $|BS\rangle=|\Psi\rangle_{AB} \otimes |\phi\rangle_C$ and two the most important representatives, the \textsl{GHZ}-state and the \textsl{W}-state:
\begin{equation}
\left|GHZ\right\rangle =\frac{\left|000\right\rangle +\left|111\right\rangle }{\sqrt{2}},\quad\left|W\right\rangle =\frac{\left|001\right\rangle +\left|010\right\rangle +\left|100\right\rangle }{\sqrt{3}}.
\nonumber
\end{equation}
Numerical results for the collectibility are compared in Table \ref{table1}.
%
We can see that the maximal and average collectibilities detect entanglement
of all three states.  The maximum value is attained for
the \textsl{GHZ}-state, $Y^{\rm max}[|GHZ\rangle]=16/64$
while $Y^{\rm max}[|W\rangle]=9/64$. As this quantity for the bi--separable state
reads $Y^{\rm max}[|BS\rangle]=4/64$
and $Y^{\rm max}[|\Psi_{\rm sep}\rangle]=1/64$,
the collectibility offers an experimentally 
accessible measure capable to distinguish the genuine
three--parties entanglement.


\begin{table}
\begin{tabular}{|c|c|c|c|}
\hline 
entanglement test  & \textsl{GHZ}-state & \textsl{W}-state & \textsl{BS}-state \tabularnewline
\hline
\hline 
minimal $Y^{min}$ & $0$ & $0$ & $0$  \tabularnewline
\hline 
maximal $Y^{max}$ & $0.250$ & $0.141$&$0.063$\tabularnewline
\hline 
average $\overline{Y}$ & $0.053$ & $0.049$&$0.021$\tabularnewline
\hline 
detection probability $P_{Y}$ & $0.807$ & $0.807$&$0.500$\tabularnewline
\hline
\end{tabular}
\caption{Comparison between the \textsl{GHZ}-state the \textit{W}-state and the bi--separable state $|BS\rangle$.
We present numerical values for the minimal, maximal and average
collectibilities and the probabilities $P_{Y}$ of entanglement detection
in a particular, random measurement. If the values of the collectibility 
are larger than $1/64\approx0.016$ then the entanglement is detected.}
\label{table1}
\end{table}

{\it Acknowledgements.-} 
It is a pleasure to thank K. Banaszek, O. G{\"u}hne, M. Ku{\'s}
and M. {\.Z}ukowski
for fruitful discussions and helpful remarks.
Financial support by the grant number  N N202 174039, N N202 090239 
and N N202 261938 of Polish Ministry of Science and Higher Education
is gratefully acknowledged.

\end{document}